%% file: MAIN.tex
\documentclass[journal]{vgtc}

\usepackage{cite}
\usepackage{amsmath,amssymb,amsfonts}
\usepackage{algorithmic}
\usepackage{textcomp}
\usepackage{float}
\usepackage{url}
\usepackage[dvipsnames]{xcolor}

\ifpdf
  \pdfoutput=1\relax                   
  \usepackage{graphicx}                
  \DeclareGraphicsExtensions{.pdf,.png,.jpg,.jpeg} 
\else
  \usepackage{graphicx}                
  \DeclareGraphicsExtensions{.eps}     
\fi%

\usepackage{microtype}                 
\PassOptionsToPackage{warn}{textcomp}  
\usepackage{textcomp}                  
\usepackage{mathptmx}                  
\usepackage{times}                     
\usepackage{cite}                      
\usepackage{tabu}                      
\usepackage{booktabs}                  

\def\BibTeX{{\rm B\kern-.05em{\sc i\kern-.025em b}\kern-.08em
    T\kern-.1667em\lower.7ex\hbox{E}\kern-.125emX}}

\title{Printmaking, Puzzles, and Studio Closets: Using Artistic Metaphors to Reimagine the User Interface for Designing Immersive Visualizations}

\author{Bridger Herman, Francesca Samsel, Annie Bares, Seth Johnson \textit{Student Member, IEEE},\\ Greg Abram, and Daniel F. Keefe, \textit{Senior Member, IEEE}}
\authorfooter{
\item
    Bridger Herman, Seth Johnson, and Daniel F. Keefe are with the Dept. of Computer Science and Engineering, University of Minnesota Twin Cities, Minneapolis, MN, United States. Emails: \{herma582, joh08230, dfk\}@umn.edu.
\item
    Annie Bares is with the University of Texas at Austin, Austin, TX, United States. Email: abares@utexas.edu.
\item
    Francesca Samsel and Greg Abram are with the Texas Advanced Computing Center, University of Texas at Austin, Austin, TX, United States. Emails: \{fsamsel, gda\}@tacc.utexas.edu.
}

\shortauthortitle{Herman \MakeLowercase{\textit{et al}}: Printmaking, Puzzles, and Studio Shelves: Using Artistic Metaphors to Reimagine the User Interface for Designing Immersive Visualizations}

\abstract{
We, as a society, need artists to help us interpret and explain science, but what does an artist's studio look like when today's science is built upon the language of large, increasingly complex data? This paper presents a data visualization design interface that lifts the barriers for artists to engage with actively studied, 3D multivariate datasets. To accomplish this, the interface must weave together the need for creative artistic processes and the challenging constraints of real-time, data-driven 3D computer graphics.   The result is an interface for a technical process, but technical in the way artistic printmaking is technical, not in the sense of computer scripting and programming. Using metaphor, computer graphics algorithms and shader program parameters are reimagined as tools in an artist's printmaking studio. These artistic metaphors and language are merged with a puzzle-piece approach to visual programming and matching iconography. Finally, artists access the interface using a web browser, making it possible to design immersive multivariate data visualizations that can be displayed in VR and AR environments using familiar drawing tablets and touch screens. We report on insights from the interdisciplinary design of the interface and early feedback from artists.}


\teaser{
    \centering
    \includegraphics[width=\linewidth]{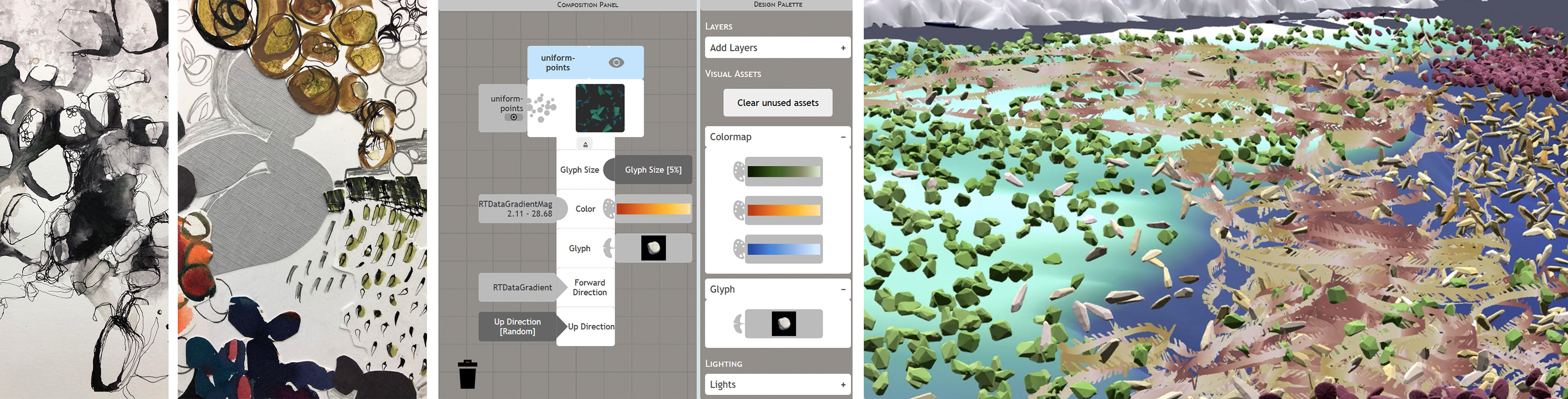}
    \caption{Here we show the progression from the artist's studio to a multivariate visualization.  Left to right: collages from Samsel's studio; the visualization design interface; a simulation of five water masses underneath the Ronne-Filchner Ice Sheet in western Antarctica.  The data is part of the DOE, BER, E3SM, developed and run by the Climate Ocean Sea-Ice Modeling group at Los Alamos National Laboratory.}
    \label{fig:teaser}
}

\begin{document}

\maketitle

\input{TEXT_Sections/1_intro}

\input{TEXT_Sections/2._PriorWork}

\input{TEXT_Sections/4_Methodolgy}

\input{TEXT_Sections/5.Results}
\input{TEXT_Sections/6._Discussion}
\input{TEXT_Sections/7._Conclusion}

\acknowledgments{This research was supported in part by the National Science Foundation (IIS-1704604 \& IIS-1704904). MPAS-Ocean simulations were conducted by Mark Petersen, Phillip Wolfram, Mathew Maltrud and Xylar Asay-Davis as part of the Energy Exascale Earth System Model (E3SM) project, funded by the U.S. Department of Energy (DOE), Office of Science, Office of Biological and Environmental Research with analyses conducted by PJW, MEM, and RXB under ARPA-E Funding Opportunity No. DE-FOA-0001726.
E3SM simulations are conducted at Argonne Leadership Computing Facility (contract DE‐AC02‐06CH11357); National Energy Research Scientific Computing Center (DE‐AC05‐00OR22725); Oak Ridge Leadership Computing Facility (DE‐AC05‐00OR22725); Argonne Nat. Lab. high‐performance computing cluster, provided by BER Earth System Modeling; and Los Alamos Nat. Lab. Institutional Computing, US DOE NNSA (DE‐AC52‐06NA25396). 
The authors would also like to thank the artists who tried out the interface, and Wingate Studio for providing the printmaking images.
}

\bibliographystyle{./abbrv-doi}
\bibliography{./MAIN}


\end{document}

%% file: TEXT_Sections/1_intro.tex
\section{Introduction}
The arts and humanities are crucial in formulating, interpreting, and expressing challenging problems and ideas, including those that are the subject of scientific inquiry like climate science, public health, ethical engagement with technology and more.  
However, our current ``data-intensive  paradigm''~\cite{jim-gray-fourth-paradigm} makes it increasingly difficult for artists, humanists, and others to engage deeply with such problems and ideas since engaging with the underlying multidimensional data increasingly requires a core background in data science or computing. 


Our team has formed a design collective, the Sculpting Visualizations
Collective. The ethos of the Collective derives from a desire to create more
enchanting visualizations, as well as improving the data-intensive tools these
visualizations are built on. Our team does this by bringing the knowledge and
experience of artists and designers into all facets of the visualization
process, from the theory behind a visualization tool to its application, design,
and the final products that it enables.

In this paper, we discuss the specific challenge of designing a user interface
for artists to create visualizations of actively studied, modern 3D datasets.
For those who wish to engage with the big, data-driven questions of our day, the
vision is to welcome that engagement in multiple modes, both as part of the
artists' own art practice and/or as part of interdisciplinary collaborative
efforts to better understand and communicate science. We want the artists
working in this space to be able to work as artists. The design user interface
elegantly mirrors the ethos of our Collective by drawing from the tradition of
printmaking. In doing so, it creates a more intuitive, enchanting way for
artists to easily and quickly iterate through many visual possibilities derived
from their work. The artist-centered design process that the interface enables
allows for artists to create visualizations that enchant users with their
surprising, yet intuitive design choices that spark imagination and curiosity.

Our design approach is grounded in an embrace of traditional, physical art processes.  We ask, what if we could make the process of providing the technical specification for a data visualization feel like it fits within the artist's studio, something that is synergistic with creative visual processes based on experimentation and ``working on the whole'' rather than the style of ``linear thinking'' and ``efficiency'' that is often more closely associated with computing and big data? 

Our technical approach makes use of the Artifact-Based Rendering (ABR) technique presented in last year's IEEE VIS technical track~\cite{johnson2019artifact}.  That technique makes it possible to render 3D visualizations using visual elements created by digitally scanning real-world artifacts (e.g., 2D paintings, drawings, ink washes, and prints; 3D clay, shaved wax, arranged objects).  The software places these visual building blocks in the 3D visual space of the data and morphs, recolors, and otherwise adapts them in response to data.  For example, working with this technique, artists can populate a virtual sea with hand-sculpted clay glyphs, coloring each one according to the ocean temperature, to help clarify for scientists five types of water masses critical to understanding ice sheet melt rates (Figure~\ref{fig:teaser}).  Artists have already used ABR to produce visualizations that exhibit a unique, decidedly hand-crafted style~\cite{samsel-visap-last-year}.  However, until now this process has required a programmer to act as a guide, helping to set up appropriate links from data objects to computer graphics render objects and to interpret the visual meaning of various parameters exposed by the computer graphics shader programs. Our technical goal is to create the interface needed to generate such a complete and correct rendering specification that the ABR engine can use to drive its computer graphics shader programs.

The interface to weave together the creative artistic process and the technical data-driven 3D rendering engine was designed and implemented over the course of approximately 8 months by an interdisciplinary collaborative team (the co-authors) who collectively bring expertise from the disciplines of visual art, environmental humanities, and computer science. The team is separated by distance, and it helped us to kick off the design project with an in-person ``hunker week,'' which was then followed up with months of implementation and iterative refinement supported by two regular team video conferences per week.  Our work began by understanding related work and then moved to building a common language, brainstorming, implementing, and iteratively refining the interface.

%% file: TEXT_Sections/2._PriorWork.tex
\section{Related Work}

Our work is inspired by the long tradition of artists engaging with and interpreting scientific data, and it contributes to literature in the interdisciplinary research areas of visual design for science and creativity support tools.

\subsection{Artists Engaging with Scientific Data}

From Leonardo da Vinci to the present VISAP program, there is a long history of artists engaging with scientific concepts and data.  The examples pictured in this paper deal specifically with climate science, a common theme in contemporary art-sci-tech work.
Artists who provide a human context to environmental issues are widespread. InfoWhelm, a recent publication by Houser, surveys contemporary art and literature addressing climate and the environment~\cite{Houser}.  Pioneered by artists including Polli, Anadol, Miebach, West, Vensa, Viegas and Wattenburg, just to name a few, the field has spurred university programs across the United States, including the University of New Mexico, the University of Oregon, Arizona State University, and UCLA ~\cite{polli2005atmospherics,Mielbach,west2015dataremix,Viegas2007,Samsel2013ArtSciTech}.  Other artists working directly from actively studied scientific data have created site-specific interactive and sculptural installations~\cite{swackhamer2017weather,singh2018orbacles}, translated topographic data to ice sculptures of glaciers that make the viewer gasp~\cite{sengal2015glacier}, and much more.   Rather than striving to produce an ``unbiased'' or ``sterile'' account of scientific data, artists often help us understand and interpret the human connection to the data, creating data visualizations that can successfully interweave both data {\em and} emotion~\cite{wang2019emotional,Carpendale-2018-art-connection}.

\subsection{Visual Design for Science}
Artists also have a rich history of contributing to the field of data visualization specifically, teaming with scientists and other data stakeholders to present data more clearly.  Cox outlined what artists can contribute to science, specifically highlighting the powerful difference that artist-designed colormaps can make as well as artists' usefulness in depicting high-dimensional spaces~\cite{cox1988using}.  Many visualization techniques have drawn inspiration from art (e.g., use of color~\cite{lum2002nprVolRend}, stippling and line rendering~\cite{lu2003stipple, almeraj2009pencilLines}, narrative forms~\cite{bach2017emerging}). Several guidelines that improve visualization clarity, engagement, and impact come directly from design theory and principles~\cite{Ware_2012,FunctionalArt}.  Numerous programs~\cite{Processing}, dashboards and entire companies have been created to enable this effort.  Most closely related to our work, powerful visualization tools, systems, and user interfaces have been designed specifically to support the role of artists in data visualization, notably~\cite{schroeder2010drawing,schroeder2015visualization,guo2011wysiwyg,bruckner2005volumeshop,keefe2008scientific, Processing}.

Since it is built upon Artifact-Based Rendering (ABR)~\cite{johnson2019artifact}, our interface supports the visualization of 3D multivariate scientific datasets using artist-created media such as glyphs, colormaps, lines, and textures.  This is a defining aspect of our approach, since this technique is already so closely tied to traditional artistic practice and leverages real-world artistic skill.  The interface also includes features found useful in other digital visualization interfaces for artists.  For example, we follow an approach similar to ColorMoves~\cite{samsel2018colormoves} to support building, editing, and tweaking custom dataset-specific colormaps in real time.

In general, the interface supports a mode of working that is fast, iterative, and visual.  Interfaces to powerful 3D visualization packages, such as ParaView~\cite{ahrens2005paraview} and VisIT~\cite{HPV:VisIt}, are primarily designed for scientists and technologists.  The workflows they provide elegantly succeed at defining a highly adjustable logical progression starting from data, moving through a processing pipeline, and finally rendering an image.  However, the approach does not necessarily align with the creative workflows of visual artists and designers.  Instead, we seek a workflow that includes a similar level of control over appearance but is more in the style of fluid, sketch-based user interfaces.  Artists have previously used sketch-based user interfaces to prototype 3D visualizations~\cite{keefe2005artistic,keefe2008scientific}, create custom illustrations of 2D fluid flows~\cite{schroeder2010drawing}, sketch free-form glyphs~\cite{schroeder2015visualization,isenberg2008interactive}, and create multi-layered animated 2D visualizations of a variety of datasets~\cite{schroeder2015visualization}.  In our interface, sketching and working with many other forms of physical media happen primarily outside of the digital interface but, in the style of Buxton's work with ``sketching user experiences''~\cite{buxton2010sketching}, working with the interface can feel like sketching in the same way that it enables a similar rapid, fluid style of visual exploration.

\subsection{Creativity Support Tools}
Tools that enable rapid visual experimentation like sketching are excellent examples of a broader category of user interfaces known as creativity support tools~\cite{shneiderman2000creating}.  Visual creativity is often aided by digital sketch pads, tablets, and other interfaces that enable artists and designers to work directly with their hands, leveraging real-world, physical skills.  Our approach is almost an extreme version of this, leveraging artists' physical real-world work by digitizing it.  Some artists have commented that this enables greater expression and personalization, since, for all the success of digital tools, there is a reason artists continue to work with physical media.  Creativity has also been linked to user interfaces that involve visual exemplars~\cite{terry2002sideViews} and fun, which is often encouraged through subtle animations and metaphors~\cite{shneiderman2004designing}.  Our work contributes to the growing interdisciplinary literature on creativity support tools with a new specific focus on artists in visualization.

%% file: TEXT_Sections/4_Methodolgy.tex
\section{The Visualization Design Interface}

We begin the discussion of the user interface design by reflecting on why this is a hard problem to solve.  There are many reasons for this, but one of the most important is the disconnect between what artists perceive as the visual object they wish to operate on and how the underlying computer graphics software organizes its data objects and rendering objects.  These various objects often have complex relationships; so, designing an interface that fits the artists' cognitive process is not as simple as making a diagram of the software organization and adding buttons and sliders to control the parameters.

\begin{figure}[!ht]
\centerline{\includegraphics[width=\columnwidth]{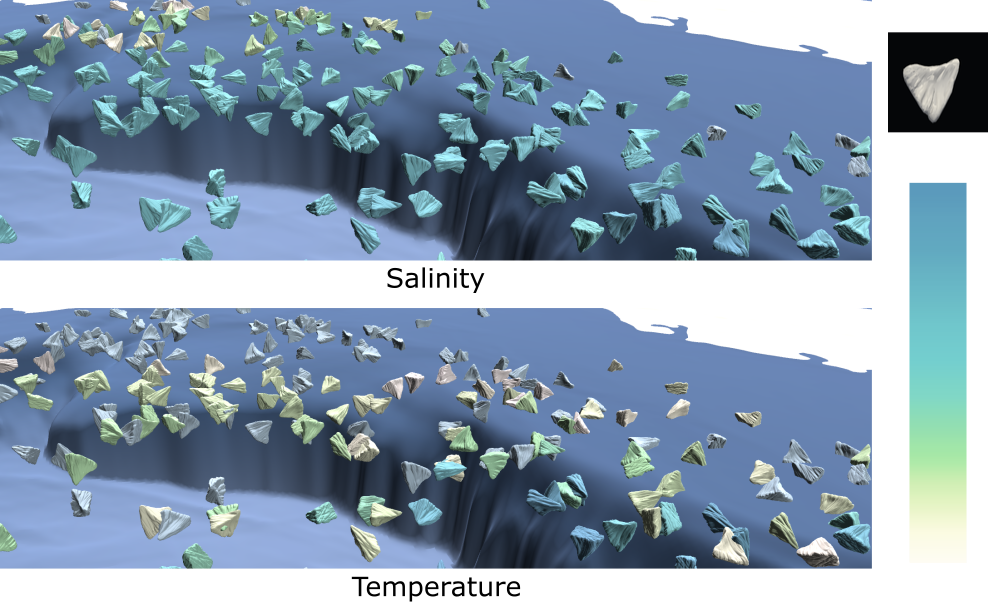}}
\caption{``Let's see what happens if we change those glyphs to show water temperature instead of salinity.''}
\label{fig:ex1}
\end{figure}

Stepping back during a design session to look at a data visualization, an artist might think, ``let's see what happens if we change those glyphs to show water temperature instead of salinity'' (Figure~\ref{fig:ex1}). 

Notice how, to the visual artist, this thought is organized conceptually around making a change to a visual object.  ``Those glyphs'' are perhaps the most visually dominant aspect of the image, and the goal is to make what seems like a small visual edit to the colormap.  
 
From the standpoint of the software, this operation actually involves three classes of data that need to be managed in concert: 1. Data Variables--temperature and salinity are both data variables; we might also call them data fields in this case because for any (x,y,z) point within the bounds of the data, we can look up the water temperature at that point. 2. A Dataset--both these variables belong to a set of data that are somehow related, in this case spatially; we might say we are visualizing ``the Gulf of Mexico dataset.''  3. Data Objects--``those glyphs'' are yet another class of data.  They exist within the same dataset, but they are not exactly data variables nor are they data fields.  We could say that they are derived from data field(s) in the sense that the glyphs are generated by sampling into another variable of the data field such as phytoplankton concentration or the nitrate concentration.  In the software, these are known by the rather nondescript term, ``data object.''  

Generally, we cannot picture data variables directly, but we can use them to modify a data object. For example, we might modify the glyph size or vary the color in response to values of a data variable from the same dataset.  This leads to rules, such as: 1) data variables and data objects both belong to datasets, and 2) you can't draw a data variable directly, but you can draw a data object and then modify some of its properties based on a data variable, as long as they belong to the same dataset.  During our ``hunker week,'' it took our team multiple days of deep discussion to believe we were finally on the same page with the nuanced relationships between data variables, data objects, and datasets, which aspects of these are precalculated and available to the user and which are not, and where data are stored -- inside a dataset, in a data object, or both.  We therefore realized a key design goal of the interface should be to naturally encode such rules and avoid the need for multi-day user training sessions.

\begin{figure}[!ht]
\centerline{\includegraphics[width=\columnwidth]{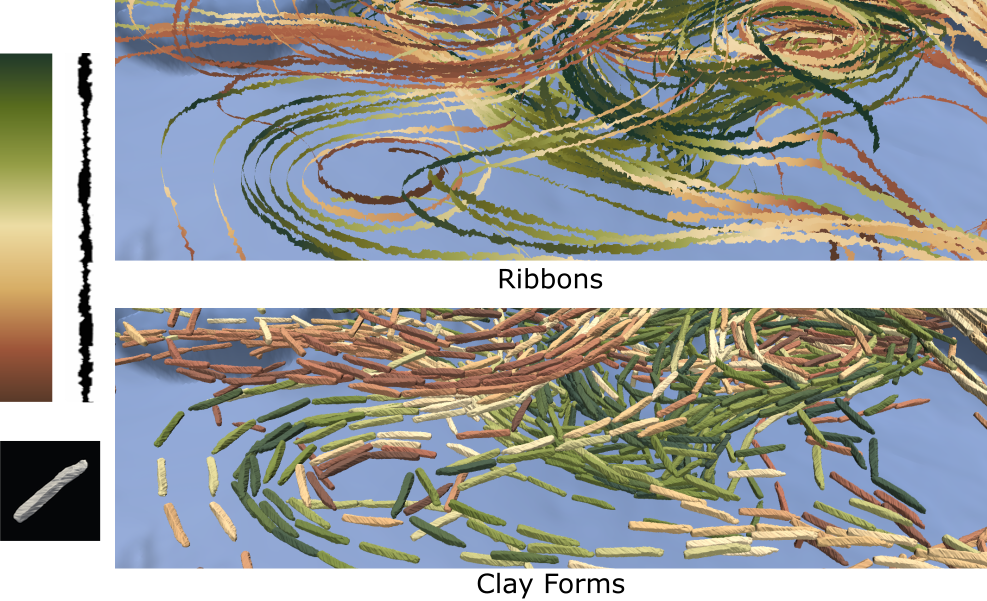}}
\caption{``Now, change the flow lines so that they are drawn using this long thin clay form I sculpted yesterday rather than the current textured ribbons.''}
\label{fig:ex2}

\end{figure}

The following request highlights another challenge, ``Now, change these flow lines so that they are drawn using this long thin clay form I sculpted yesterday rather than the current textured ribbons'' (Figure~\ref{fig:ex2}).

\begin{figure}[t]
\centerline{\includegraphics[width=\columnwidth]{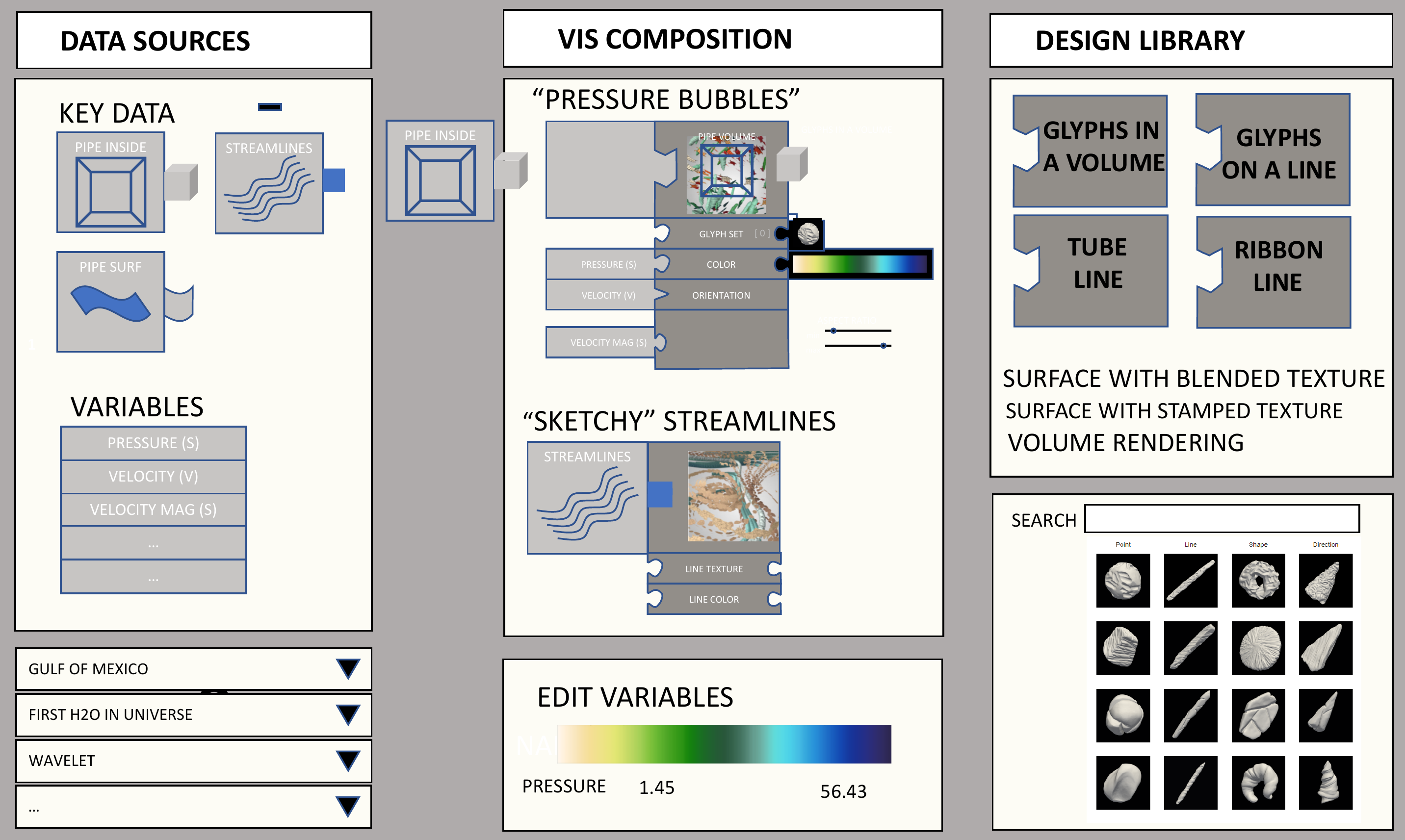}}
\caption{Digital mockup of the three-panel interface.  The center Composition Panel is where artists link together data, stored on the left panel, and visual elements, stored on the right.}
\label{fig:GUImockup}
\end{figure}

Again, from the artist's standpoint, this request is completely natural.  It operates on the visual element of the ``flow lines.''  On the surface, it does not ask to change that object in any fundamental way--they are still the same lines.  The request is just to change the visual representation.  Unfortunately, to the computer graphics programmer, this must be interpreted as a big change because the efficient approach to drawing these lines is completely different in the two cases.  The differences, even requiring different triangle mesh geometries and different shader programs, are significant enough that most graphics programmers would take the approach, ``OK, if you want to make that change, then I will just delete the old ribbon render objects I created for the lines earlier and start over, making a new set of lines that use my instanced mesh shader program to create something that looks visually like a line but is really many copies of your clay mesh.''  Creating a user interface that embraces the language and creative visual design processes of artists while also preserving the full power of the computer graphics and efficiencies needed to render large datasets in VR is a serious challenge.

\subsection{The Language of Printmaking}

One of the joyful struggles of interdisciplinary collaboration is breaking down the traditions, processes, and assumptions of our various disciplines in order to find the intersection points that we {\em know} are there but can be so difficult to articulate clearly.  We often use metaphor to assist with building a common language in such conversations.  So, during the ``hunker week'' that kicked off our interface development effort, we searched for metaphors that might help to translate the most complex requirements of the 3D graphics data rendering engine into a language that fits with artistic traditions.

Printmaking emerged as one of the most useful metaphors.  In intaglio printmaking, artists first create a design by carving or etching into a matrix, such as a metal plate.  Ink is applied to the plate, filling in the recessed design.  Then, the print is pulled (the design is transferred to paper) by running the paper and plate through a press.  A single transfer in this style is called an impression, but prints routinely combine several layers of impressions from multiple different plates.  An edition of identical prints can be produced by repeating the process with the same inks, plates, and ordering, or, an infinite variety of new prints can be created by combining plates in new ways, inking them differently, or even adjusting pressure on the press.  There are many additional variations to the process including that wood, metal, stone, and linoleum can all be used as matrices.  However, one amusing constant seems to be that if you ask a printmaker, ``do you have any old plates in your studio?''  You will often get a smile and the answer, ``yes, how did you know my closet is absolutely overflowing with metal plates, wood blocks, and ink!''

There are some useful connections between printmaking and data visualization.  Like a library of reusable computer graphics algorithms that can draw points, lines, or surfaces in different colors and locations based on the data sent to them, that closet full of plates provides a reusable collection of design elements that can be reinked and rearranged on the page in countless ways.  Waiting in the closet until they are needed, these plates (rendering algorithms) are only brought to life when they are loaded with ink (data).  It is true that printmaking notably departs from 3D computer graphics in that the result is typically a 2D image; however, printmaking is also the one traditional, physical art form where the concept of building up a complete composition from a series of separate layers is absolutely obvious.  In fact, the technical steps required to set up and pull each layer can be quite complex and time consuming.  So, many printmakers naturally think in terms of layers and procedures that are quite reminiscent of the way multiple data objects (streamlines, surfaces, points) are sent in serial through the computer graphics pipeline and ultimately superimposed to form a complete 3D visualization.

\subsection{Early Ideation and Sketching}

The concept of reusable plates that can be inked with data forms the foundation for the interface.  When combined with data, each plate forms a ``data impression,'' a visual layer that becomes part of the overall composition.

Another high-level concept is the notion of data and visual art/design meeting
in the middle to create a visualization.  Figure~\ref{fig:GUImockup} shows an
early design mockup exploring this concept.  Notice the 3-panel design with the
``Data Sources'' on the left, the ``Design Library'' on the right, and the ``Composition'' panel in the middle.  Adding data impressions to the composition
requires bringing together elements from both the left and the right.  Most
traditional 3D data visualization pipelines start from the data and proceed in a
linear fashion through a pipeline until an image is rendered at the end.  In an
effort to better support design that places a priority of visual decisions, this
interface makes it possible to start with the design, pulling colors, forms, and
textures into the composition before linking them with data.

The links between data and visuals could be made by drawing lines between input and output ports to form a pipeline in the style of Data Explorer~\cite{greg-ibm-data-explorer}, but we designed an alternative inspired by the block-based visual programming tools often used to teach programming~\cite{pasternak2017blockly,resnick2009scratch}.  This puzzle piece metaphor has the great advantage of visually encoding the difficult-to-explain rules mentioned in the earlier discussion of ``why designing this interface is a hard problem.''

The puzzle connectors are designed to be abstract enough to work as interface elements, but specific enough to evoke the sense of the item they are representing.
For example, the point data connector looks like bubbles coming out of the puzzle piece, and the colormap connector looks like a painter's palette.
Also worth mentioning are the action icons located throughout the interface which use the Material Design icon pack~\cite{materialDesign}, which is a popular choice among smartphone and web apps alike.

\subsection{Data Impressions}

 \begin{figure*}[htbp]
    \centering
    \includegraphics[width=\textwidth]{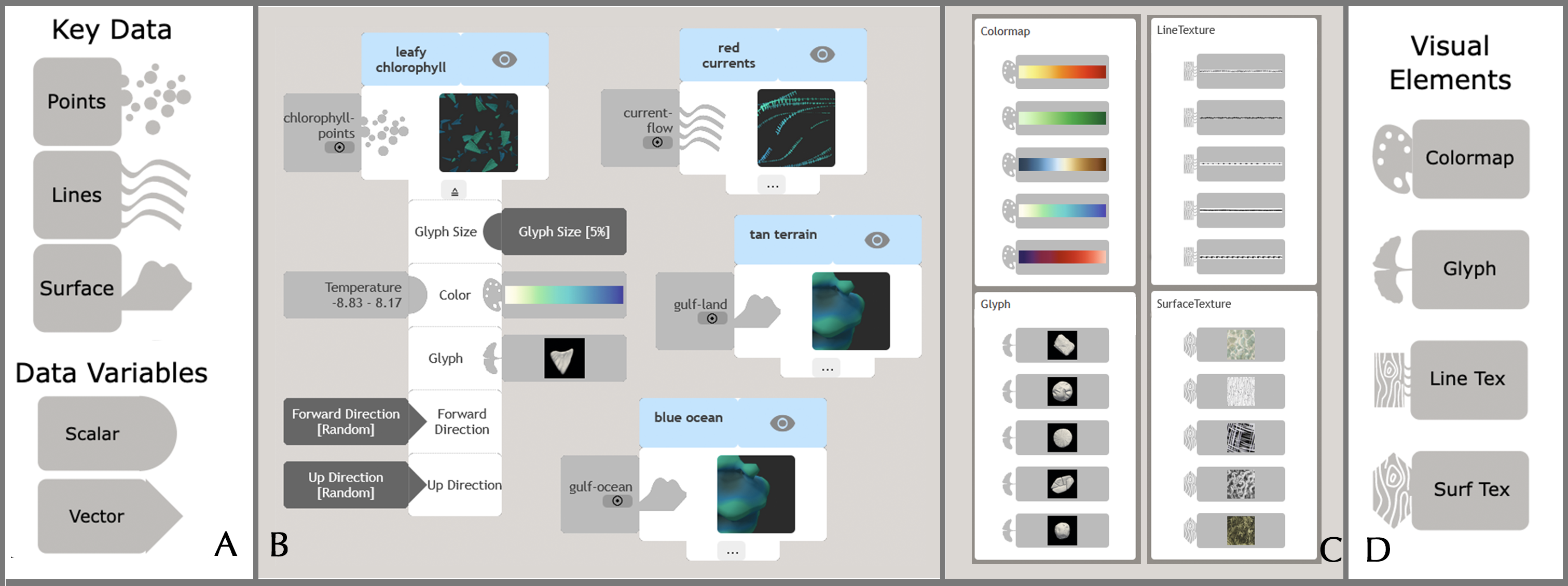}
    \caption{(A and D): The iconography used in the interface, showing the types of Key Data, Data Variables, and Visual Elements available to the artist in the interface. (B): Example of a glyph plate ``Leafy Chlorophyll'' which is registered with the ``chlorophyll-points'' density-based point sampling and has the ``Temperature'' scalar data variable encoded with a colormap from the design palette, as well as a ``drum'' glyph. The dark gray slots indicate entries that have not yet been assigned by the artist and remain at the default values shown in brackets. Other types of plates (ribbons and surfaces) are also shown. (C): 20 examples of Visual Elements available to the artist in the interface, including Colormaps, Glyphs, Lines, and Textures.}
    \label{fig:icons_and_plates}
 \end{figure*}


The complete, current visual language of plates, data, and visual elements is shown in Figure~\ref{fig:icons_and_plates}.  Let us describe how artists work with these building blocks to design a visualization composed of multiple data impressions.




There are several ways to create a data impression, but a typical approach starts with selecting a plate from the design palette in the right panel, also known as the ``studio closet.''  This collection of plates is expandable.  Whenever a new data-driven rendering algorithm is added to ABR, we add a new plate type.  The pattern on each plate is an example of the visual style it can create, but this is just one example.  The plate will produce something different depending on how it is inked.

Since the rendering is 3D, the plate needs to know where its pattern should be
applied in space. In the underlying technical system, this information comes
from the ``data object'' discussed earlier, which is, in our experience, one of
the most challenging technical aspects to explain. Our interface addresses this
by, again, using metaphor. Printmakers are familiar with the need to align or
register a pattern, and they often use registration or ``key'' plates to
accomplish this. Thus, the interface presents data objects as ``key'' data.

Key data are necessary. There is no way to draw data variables without providing
some spatial registration, so every plate includes at least one puzzle-piece
slot for key data. Notice, also, that the slot types vary based on the different
icons for registering the plate's pattern to 3D locations in space (points), 3D
curves (lines), and forms (surfaces). Figure~\ref{fig:icons_and_plates} shows
just the plates available in our current implementation. Going forward, we
expect our team will always include computer graphics researchers and artists
working together to develop new plates and corresponding graphics algorithms.

A specific example of a ``Leafy Chlorophyll'' data impression designed by an artist is shown in Figure~\ref{fig:icons_and_plates}.  As we see from the shape of the puzzle piece, this plate only works when registered to point-style key data.  In the Gulf of Mexico Biogeochemistry dataset pictured in several figures of this paper, there are multiple options associated with this type of key data, including ``Chlorophyll Concentration'' and ``Nitrate Concentration.''  In this case, the artist has registered the plate to ``Chlorophyll Concentration.''  We often use ``Concentration'' to signify a density-based sampling, which means the pattern will be sparse in areas of low concentration and dense in areas of high concentration.

Beyond the spatial registration provided by key data, all of the plates we have created to date also include additional settings to further customize the data impression.  The interface presents these in a collapsible list attached to the bottom of each plate.  When a plate is placed in the center Composition Panel, the list automatically opens up to show all available options, as shown in the ``Leafy Chlorophyll'' example.

All of the settings in this list are optional, and we require all plates to have
reasonable defaults. So, as soon as the plate is registered with key data, a 3D
visual will appear in ABR, which is usually set up to run side-by-side on a
second monitor and/or in an attached VR/AR headset. This enables the artist to
explore the visual results in real-time 3D and react with visual changes.
Several settings have been adjusted in the ``Leafy Chlorophyll'' example by
attaching data variable puzzle pieces from the left data panel and visual design
puzzle pieces from the right design panel.

Returning to some of the technical challenges that must be overcome in the interface design, note the separation in this interface between key data and data variables.  With the settings shown in Figure~\ref{fig:icons_and_plates}, the ``Leafy Chlorophyll'' data impression produces a visualization like the one in Figure~\ref{fig:ex1} {\em Temperature}.  Making the design change described in that Before--After example is trivial.  The artist must only replace the ``Salinity'' puzzle piece in the ``Color Variable'' slot with the ``Temperature'' puzzle piece from the scalar variables section of the data palette on the left panel of the interface.

Beyond this quick switch to a different data variable it is also possible for the artist to decide that the way that pattern is distributed in space is not working.  Perhaps, the combination of leafy glyphs works perfectly with the background color scheme and it makes sense to include these somewhere in the composition, but the chlorophyll concentration data happens to be very unevenly distributed and the leafy glyphs are all clustering together in a way that is not very readable.  It is possible in this case to keep all of the existing data and visual settings the same but swap in different key data, ``Nitrate Concentration,'' for example.  This will have the effect of retaining all of the plate's visual style but re-registering the pattern to a different 3D spatial distribution.

Let us consider the other Before--After picture mentioned earlier.  Recall that the design edit illustrated in Figure~\ref{fig:ex2} is a drastic change from the standpoint of the computer graphics algorithm that should be used, even though artists think of it as a visual change to the same set of lines.  This example requires more manipulation of the interface to achieve, but all of the underlying complexity is hidden behind the metaphors and iconography.  The {\em Ribbons} image was created using a textured ribbon plate where artists can provide several styles of texture, including one used here to give the ribbons their patterned edge.  The ribbons are also colored according to data.  To use a sculpted clay form to depict the lines instead, the artist must first notice that the ribbon plate does not include a slot for glyph key data--it is impossible to ink this plate with glyphs because the underlying software approach to the two styles of line is so different.  The solution is to ``go back to the closet'' and find a different plate that is more suited to glyphs.  Once they place this new plate in the composition panel, they can move all the important repeated elements (key data, data variables, color) from the original plate to the new one.

The composition panel holds all of the data impressions created for the visualization.  Artists often reposition these within the panel to organize the space, and the panel itself can be panned if the artist has more layers than they have screen space.  Drawing inspiration from digital image manipulation software, each data impression also includes buttons to hide, collapse, expand, or delete the impression.
When an artist saves their work, the placement of each layer in the composition panel persists between sessions.

\subsection{Importing and Editing Visual Assets}

Artists seldom work in a vacuum, and the interface embraces this concept by
enabling artists to incorporate any visual elements stored in the public
Sculpting Vis Library~\cite{SculptingVisLibrary} in a way that feels magical.
The Sculpting Vis library is simply loaded in one web browser window while the
design interface runs in another.  Then, any of the glyphs, colormaps, and
textures in the library can included by simply clicking on their thumbnail images in the
library window and dragging these into the design interface browser window.
This automatically adds them to the current working palette (right panel) and
triggers the connected instance of ABR to download the original raw 3D model
files, image data, etc. so that these elements may be used for 3D computer
graphics rendering.

The library itself is rapidly expanding and currently contains a selection of glyphs, colormaps, lines, and textures curated by artists in the Sculpting Vis Collective.  Individuals may also upload new visual elements to the library via the ABR applets~\cite{johnson2019artifact}.

Rather than working to create generic, generalizable color maps, glyphs, and textures, one of the benefits of this design interface is that it makes it increasingly practical to create data-specific visualizations, that is, visualizations that include color maps, glyphs, and other elements that are fine-tuned for the particular data at hand.  To this end, the interface includes a data-specific colormap editor inspired by the powerful ColorMoves tool~\cite{samsel2018colormoves}.
Double-clicking the color map's thumbnail icon in the interface launches the colormap editor, which shows the colormap on top of the data histogram for the variable that the colormap is attached to.  This allows the artist to tune the colormap relative to the actual data values.  The interface includes features to add, subtract, and adjust the colormap's control points.




\begin{figure*}[htbp]
    \centering
    \includegraphics[width=\textwidth]{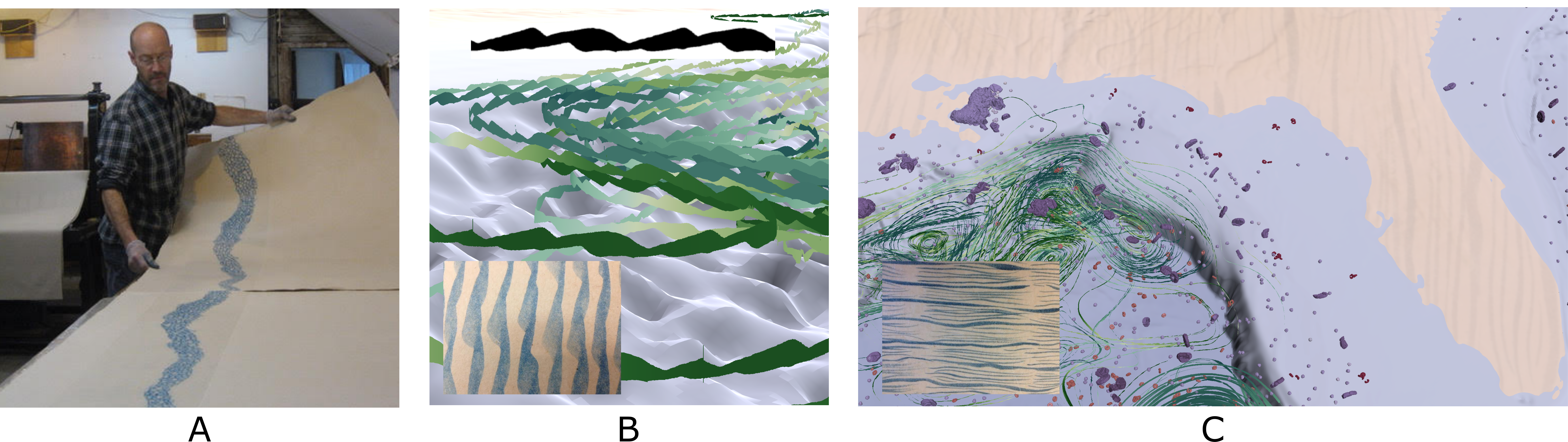}
    \caption{Deborra Stewart-Pettengill works with master printmakers at Wingate Studio (A) to realize her chine-coll\'{e} designs. She commonly works with patterns, which have been digitized to form streamlines in the Gulf of Mexico visualization (B) and a texture on the land (C). Image (A) Copyright 2020 Wingate Studio; used with permission.}
    \label{fig:artist1_gulf}
\end{figure*}

\subsection{Implementation}

To support a multitude of devices including desktop and laptop computers, tablets, and large-format touch screen displays, the interface is built for web browsers using a combination of Django, JavaScript, jQuery UI, and other libraries like Data Driven Documents~\cite{zhu2013d3}.  
The puzzle piece connections are implemented with snapping so that a connection is made whenever a correct puzzle piece is dropped near, even if not precisely on, a valid slot.  If the slot is already occupied with another piece, it is replaced, and if the piece does not match the slot, the connection is refused.  In this way, the implementation makes it impossible to drop a ``Glyph'' visual element into a ``Colormap'' slot in a data impression, and it is impossible to drop a scalar variable into a vector variable slot.

To support diverse applications, the system is implemented using a modular structure, where the web-based design interface, the ABR 3D rendering engine, and the data engine are three distinct sub-systems that connect to each other using network sockets.  
This means, for example, that the data can be hosted on a supercomputer, the graphics can be rendered on a machine with a powerful graphics card, and a designer can craft a visualization on a tablet while interactively monitoring their design modifications on a laptop with a remote viewer.
This opens up possibilities for more artists to become involved in the visualization design process by drastically reducing the hardware requirements for building visualizations of large scientific datasets.

We have extended ParaView~\cite{ahrens2005paraview} to act as a data server within this framework.  To create the figures shown in this paper, we used ParaView to prepare datasets from two supercomputer climate simulations actively studied by our collaborators: Biogeochemistry in the Gulf of Mexico~\cite{wolfram2015diagnosing,Ringler_ea13om,Petersen_ea15om} and Sea Ice Climate data~\cite{petersen2019antarctic}. The design interface does not support data wrangling itself; filtering data, deriving new data variables, and similar operations can be done in ParaView before or during a design session.  This is one limitation in the sense that we do not expect artists to have knowledge to do this level of data wrangling.  However, after data have been loaded into ParaView once and saved, the design interface makes reloading these data for design work simple through its Load and Save functionality.

The three modules are connected as follows: The geometric representation of the
data is sent from the data server (ParaView) to the rendering engine (ABR),
which pre-processes and optimizes the data for rendering. The ABR engine
connects to a Python server which delivers the interface to the artist. As the
artist makes changes to the visual styles on the interface, messages are sent
via WebSocket to the engine, which updates the encodings that are rendered
accordingly. Graphics from the engine can optionally be rendered to a depth
texture~\cite{luke2002semotus}, and sent to any connected remote viewers. VR
headsets can also be connected directly to the engine, which renders the
visualization graphics at 120+ frames per second. The scale of the
visualizations in VR is initialized to a table-scale default, but can be
adjusted via bimanual interactions from the user. Graphics can also be easily
exported as PNG images, which is useful for artists using an ABR-created
visualization as a part of a transmedia piece.

%% file: TEXT_Sections/5.Results.tex
\section{Insights from Artists}

This section discusses the experience and insights from artists on our team who have had a chance to work with the interface in detail as well as feedback artists who have just started the process of designing visualizations with the interface.  We begin by describing design sessions we facilitated with two practicing artists as an introduction to the tool.

We cast these two introduction / design sessions, simply, as a chance to work creatively with a dataset scientists are currently using to understand the biogeochemistry of the Gulf of Mexico~\cite{wolfram2015diagnosing}. These data include surface layers for the ocean and land, streamlines for the ocean currents representing the direction of eddy curvature (clockwise or anticlockwise), and concentrations for nitrates and chlorophyll.  The sessions were conducted remotely so as to respect the social distancing necessitated by the ongoing pandemic.  The software ran on a local computer with the artists connecting remotely over a video conferencing link.  In one case, that link supported having the artist take over control of the local computer and use the interface with her own mouse.  In the other, we acted as the artist's hands on the interface, sharing the screen over video and following her directions.

\subsection{Vis Design Session 1}

The first artist, Deborra Stewart-Pettengill, works with many forms of traditional artistic media and had no experience working with 3D scientific data prior to this session. Her current art processes include printmaking with chine-coll\'{e}, an intricate technique that involves adding color and form to a print by using the press to apply shapes of thin cutout paper to the print.  She was interested in experimenting with her abstracted natural and organic forms on a data visualization after she was contacted by a member of the Sculpting Vis Collective. 

After some introduction to the project and discussion to find common ground, we realized we were all curious to see how the line quality she and the master printmakers at Wingate Studio achieve with chine-coll\'{e} (Figure~\ref{fig:artist1_gulf}A) might translate into a digital data visualization.  So, she digitized some of this work using her smartphone camera, and we used the ABR's Infinite Line and Texture Mapper applets~\cite{johnson2019artifact} to prepare the results for attaching to data.  

``I'd like to start with something neutral in the terrain,'' Stewart-Pettengill stated at the onset of our session; together, we assigned a generic brown colormap and left it for the time being.  Later, toward the end of the session, she returned to this color; ``the one thing I want to change is the color of the land.''
The colormap was modified to tease out the most varied part of the elevation data -- ``That's looking a lot better, the peachiness of it looks good with the blue-violet water.'' 
Stewart-Pettengill was enthusiastic about including her own chine-coll\'{e} work
-- when she first saw the 8 chine-coll\'{e} textures in her palette, she immediately recognized them: ``Oh
yeah, there are my strokes! Cool!''  Later, she applied these to the streamlines depicting ocean
currents and experimented with adjusting line width and coloration
(Figure~\ref{fig:artist1_gulf}B,C).
In fact, she repeatedly came back to the streamline pattern and colormap throughout the process, particularly once she began assigning encodings to the chlorophyll and nitrate glyph layers.  

We will return to the theme of the importance of bringing one's own work from the studio into the new project of visualizing data.  This emerged as a consistent theme.

Another important lesson from the introductory session with Stewart-Pettingill was the need for visual experimentation.  She is ``just so used to experimenting with things'' as she works, and reflected that the interface enabled a similar process during her session. To provide a sense of the scope of experimentation that is possible, we noticed that in addition to the 8 line textures mentioned earlier, she also worked with over 10 glyphs and more than 20 colormaps for the ocean current streamlines, the nitrates, and the ocean floor during this first-ever design session.





\begin{figure}[tbh]
\centering
\includegraphics[width=\columnwidth]{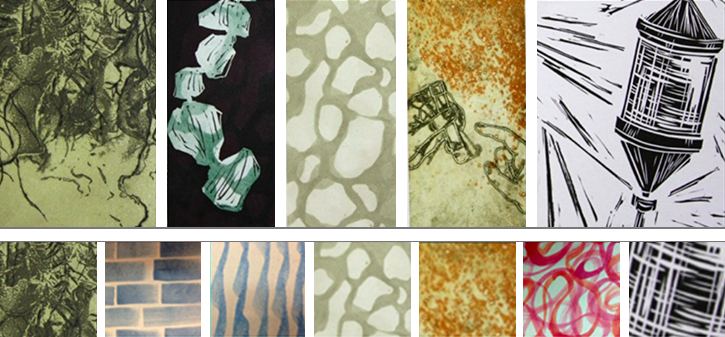}
\caption{The top image shows chine-coll\'{e} etchings by Samsel, from a series entitled Osmosis. Details from the etching are converted into textures usable in ABR visualizations, shown in the bottom row alongside textures from Stewart-Pettengill.}
\label{fig:texture}
\end{figure}

\subsection{Vis Design Session 2}

The second artist, Stephanie Zeller, had prior experience working with 3D scientific visualizations including custom colormap creation and has worked with the Sculpting Vis Collective previously. Zeller is trained in traditional media with a focus on painting. Her work of late examines the viewer's relationship with the digital environment through traditional media, especially in terms of data consumption and analysis. Zeller is also a freelance writer and uses a breadth of digital software to augment her work in data journalism through both digital illustration and information visualization.

Of her session, Zeller remarked ``As I worked, I was coming first from an
artistic angle, closely followed by a scientific angle. I wanted it to be both
aesthetically appealing and to also maintain the ability to distinguish between
variables, which requires a fair amount of stepping back and moving in again.''
She spent a short time experimenting with colormaps for the ocean floor once a texture was applied, then moved on to working on the terrain colormap.  She clearly judged the terrain as critical to the composition, spending more than 20 minutes to fine-tune a brown/tan hued colormap to work well specifically for these data.  She later decided to take the design in a new direction, shifting to a green/white colormap to achieve suitable contrast with the rest of the visualization elements (particularly the streamlines) and to not overpower the main subject, which is the visualization in the ocean.

The design process was clearly iterative, with each design decision coming as a
reaction to what was seen.  At one point, Zeller decided that there was ``too
much green'' in the visualization, and swapped out the old olive chlorophyll glyphs
for a blue instead.  Later, she returned to the colormaps for the land and the
streamlines, making fine adjustments to achieve a balance between them.

\begin{figure}[t]
    \centering
    \includegraphics[width=\columnwidth]{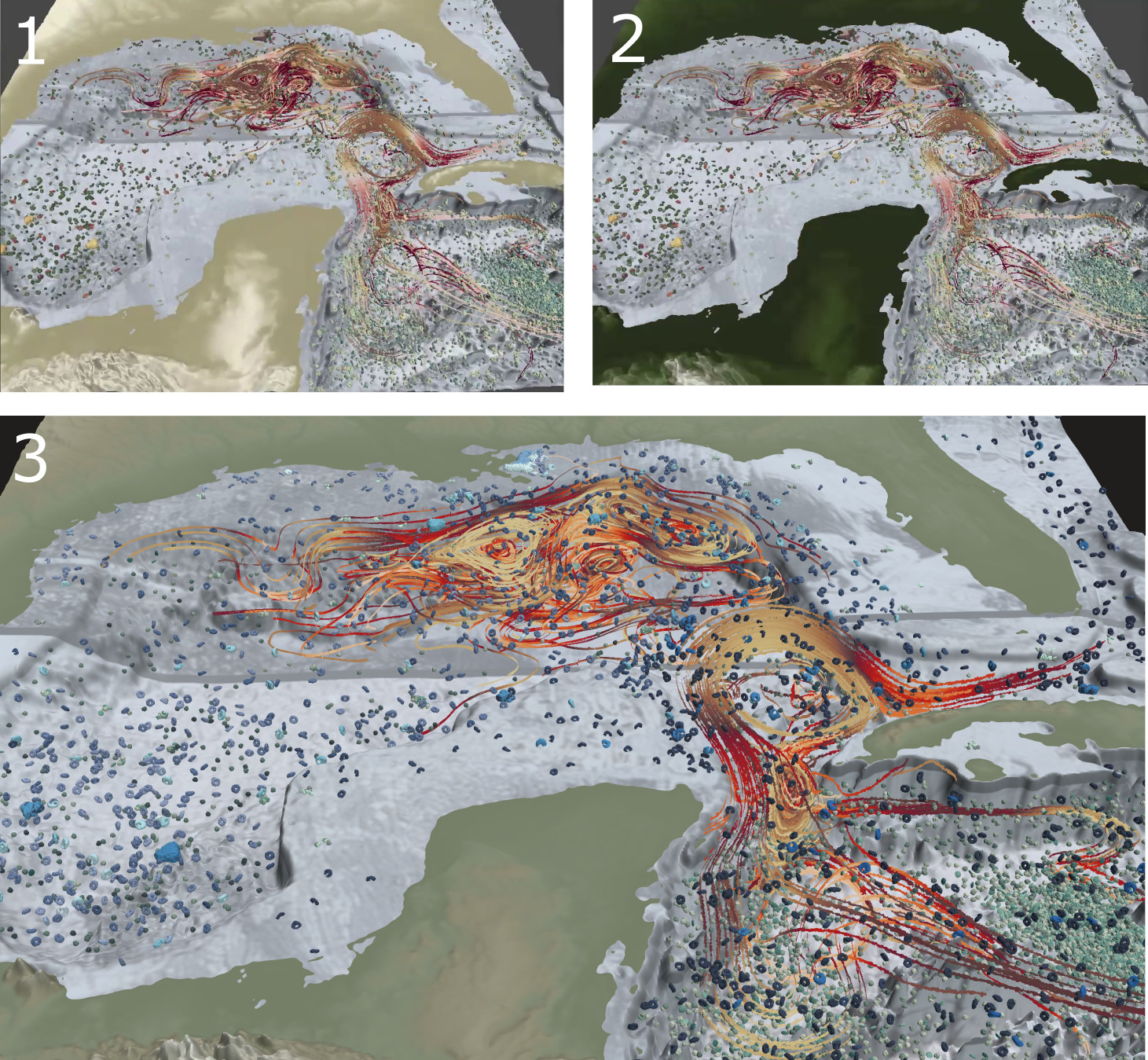}
    \caption{Stephanie Zeller's process for designing a visualization of the Gulf of Mexico. (1) shows an early step in the visualization, (2) includes progress on the streamlines and terrain colormap, and (3) shows the final result including several custom colormaps designed by Zeller.}
    \label{fig:artist2_gulf}
\end{figure}

Similar to Stewart-Pettengill, Zeller was also keen to incorporate her own artwork in the visualization; in her case through color.  During this introductory session, she worked with more than 50 different colormaps, editing 18 of them relative to the data histogram, and creating 2 more from scratch.  She felt that many of the colormaps given in the curated library didn't resonate with her style and created new ones adapted from her own work in data journalism, which can be seen on the ocean current streamlines (Figure~\ref{fig:artist2_gulf}).


In her own paintings, Zeller often builds very large surfaces (8 feet or more in
length), and works with details up close.
Every so often, she steps far enough back to see the entire piece and understand
the color relationships in the part she just worked on, and make an informed
next move in the part.
Building visualizations with the interface was appealing to her because she felt
she could address design concerns from up close and far away using her familiar
processes.
This was reflected in her usage of the interface -- when choosing colors and
glyphs, she started with detail-oriented view, quickly moved to a more global
perspective, then made changes in the visualization design based on how the
color interactions between variables both close up and far away.
Zeller felt that for similar reasons, it was intuitive to design for both a
close-up 3D perspective and a bird's-eye 2D perspective using the interface.



\begin{figure*}[tbh]
\centering
\includegraphics[width=\textwidth]{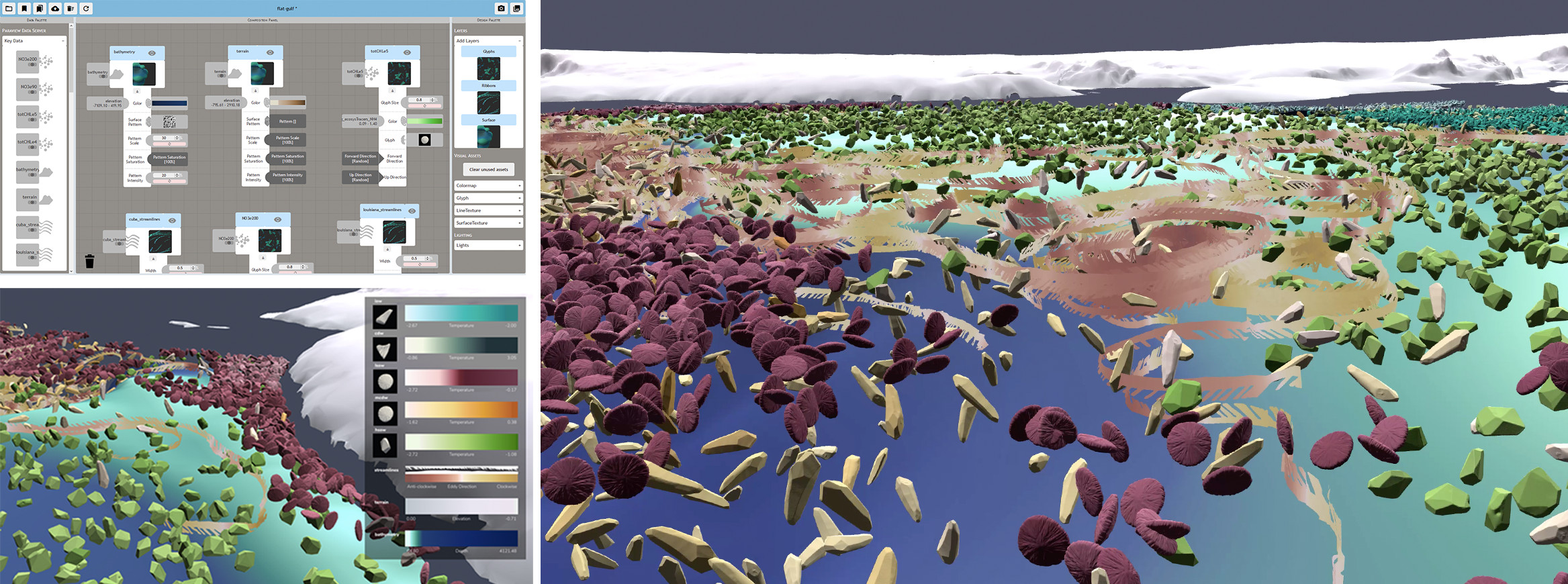}
\caption{This visualization of water masses underneath the Ronne-Filchner Ice Sheet shows: five ocean masses, their locations and temperature; two directions of eddy flow; the ocean floor depth; and the topography of the Antarctic. Here we are illustrating the power of including artists in the process. In order to render seven overlapping variables in a 3D simulation, the visualization uses distinguishable glyphs and hues. The visualization shown here provides scientific value, as scientists have not previously seen the movement of these water masses which are critical to predicting the melt rate of the ice on the underneath side of the Antarctic ice sheets. Data - E3SM, BER, DOE.}
\label{fig:science_connection}
\end{figure*}

\begin{figure}[h!]
\centering
\includegraphics[width=\columnwidth]{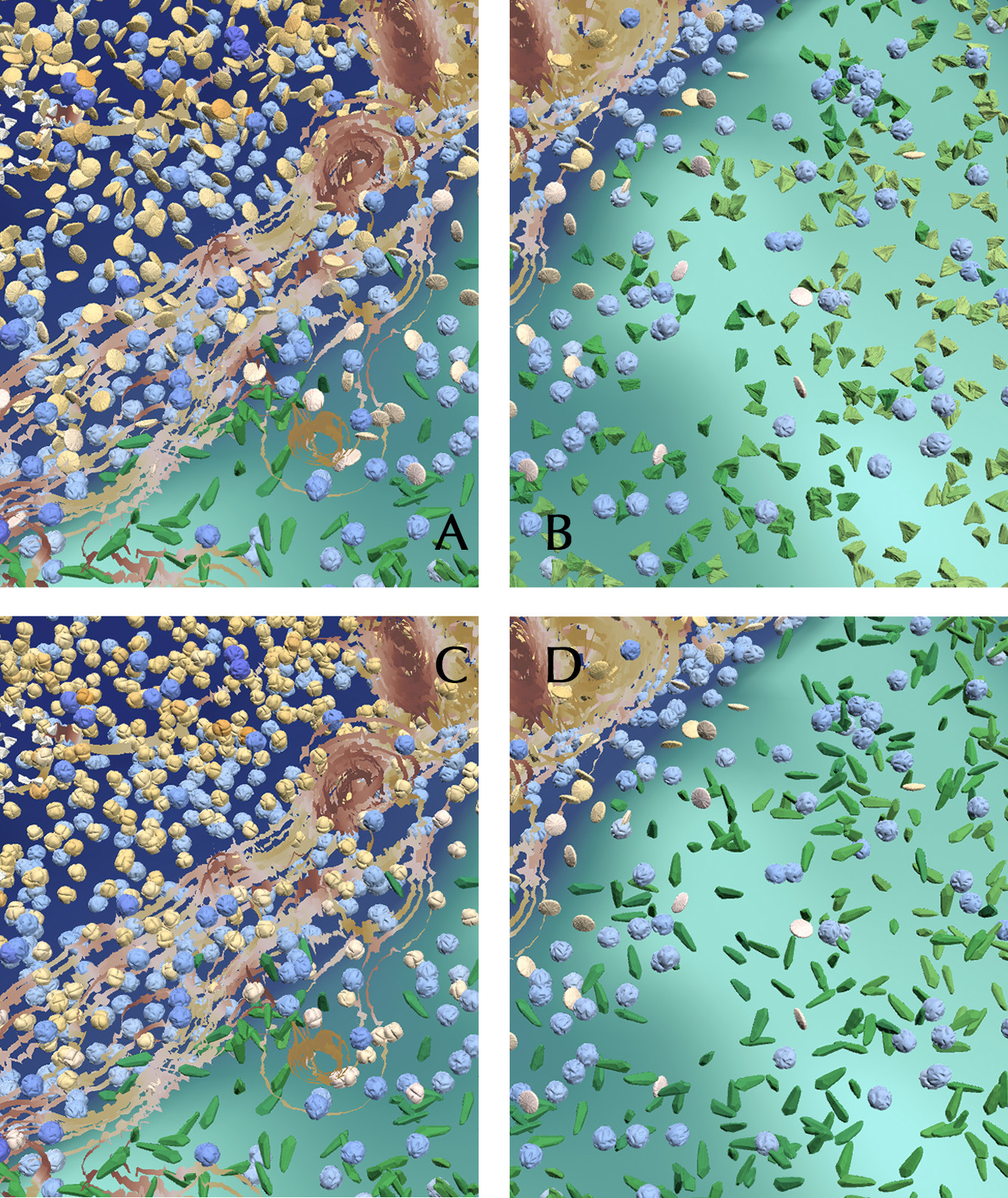}
\caption{Detail in the glyph selection process. The yellow glyphs in the left column (A and C) compare a disk verses a spherical glyph. On the right, the green glyphs compare a triangular shape (B) verse an elongated form (D).  While these are subtle shifts, they provide critical contrast when working on large complex data. }
\label{fig:glyphs}
\end{figure}

%% file: TEXT_Sections/6._Discussion.tex
\subsection{Interpretation and Extended Use}

\begin{figure}[tbh]
\centerline{\includegraphics[width=\columnwidth]{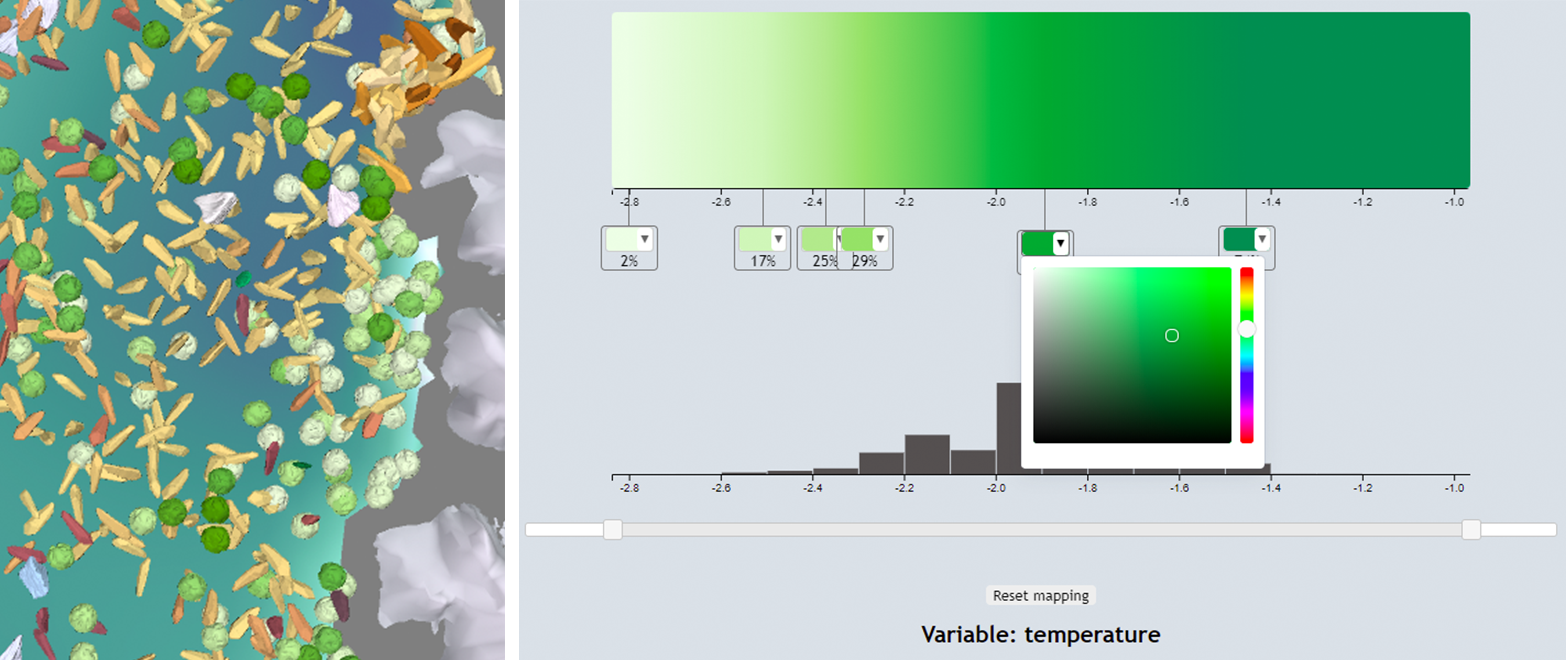}}
\caption{The includes the ability for artists to upload and or create new colormaps via the Color Loom applet using the ABR technique.  While this provides limitless options, the artists felt strongly that they needed to be able to control the distribution of the hues across the data in order to create the contrast and visual distinction with in the imagery.  Thus an internal colormapping tool was added to the interface. On the right is the color interface enabling on to control hue distributions. }
\label{fig:colormap-interface}
\end{figure}


\begin{figure}[tbh]
\centering
\includegraphics[width=\columnwidth]{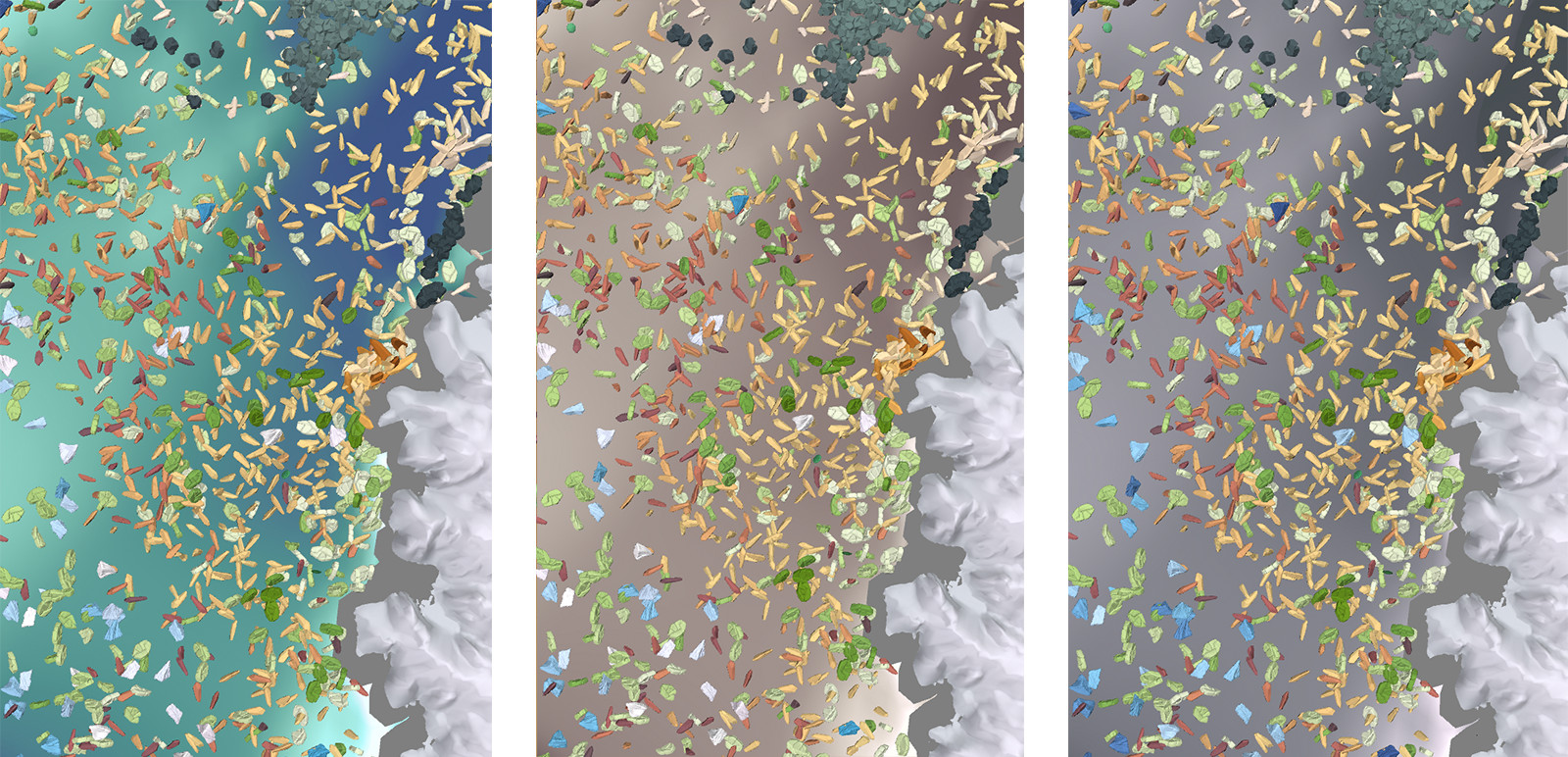}
\caption{Intuition might dictate that the water in an Antarctic visualization should be blue (left). However, in practice with complex multivariate data such as the six water masses mixing shown in this figure, using blue as the ocean can quickly become overwhelming. One solution is to try a more neutral color (middle and right). The subtle differences between the middle and right images is something easily accomplished in the ABR interface, enabling fine-tuned control of visualizations.}
\label{fig:background}
\end{figure}

Francesca Samsel, an artist in the Sculpting Vis Collective is a co-designer of the interface and has worked with it in various stages of development to create works like those pictured in Figure~\ref{fig:teaser} and \ref{fig:science_connection}, which build so clearly upon her art practice, for example, the form, line quality, metaphor, and color of the Osmosis series (Figure~\ref{fig:texture}).

In this section, we report both on Samsel's own observations working with the interface and her reflections on the feedback received from other artists.  Together, these coalesce around three major themes.

\subsubsection{Scope and fidelity of visual vocabulary}


The visual variation exhibited by artists was significant because the workflow
presented by the interface allows each individual to contribute their artistic
vision to the visualization design.  In many cases, this is accomplished by
enabling artists to utilize elements that they have previously created, such as
chine coll\'{e} and custom colors discussed earlier.  Samsel shows how the same
can be accomplished with hand-sculpted clay forms to create custom glyph
vocabularies (Figure~\ref{fig:glyphs}).  Relative to typical scientific
visualization software, the scope of visual variation that is possible is much
larger, similar to working in a studio.  So, subtle variations in visualization
design can be explored
to discover combinations that work together.

Additionally, since the interface is focused on enabling an interactive design
process, artists can make minute changes to the visual inputs, and often these
have a profound effect on the resulting visualization.
Figure~\ref{fig:background} demonstrates how an artist with the ability to fine
tune the color can apply color contrast principals to impact the visualization
for clarity.

We see it as a mark of a good interface that this fidelity to fine-tune the visualization correlates with artistic skill.  The ability to render an object without thinking is the root of visual invention.  This is the means via which artists push their work/vocabulary forward. The idea is to leave the critic outside the room and trust in the process.  The key to the arts is to gain the technical skill so that your hands flow freely,  then when you enter the studio to turn off the linear or rational decision making process and follow the work itself as it guides you to the next step.









\subsubsection{Rapid iteration and stimulation of artistic imagination}

The scope and fidelity of the visual vocabulary combined with the speed at which one can explore visual alternatives seems to lead to a tool that stimulates the artistic imagination.  ``This is fun!'' -- both of the artists seeing the interface for the first time expressed enjoyment in the process.  Artists can get into the creative zone, that special mental place where artistic magic happens, even while they are working with 3D multivariate scientific datasets.


It is necessary that any tool living up to an artist's specifications must easily support iteration in order to facilitate work in the creative zone. While Figures~\ref{fig:glyphs},~\ref{fig:colormap-interface}, and ~\ref{fig:background} demonstrate the impact of shape and color on our ability to distinguish between variables, Figure~\ref{fig:science_connection} demonstrates that engagement is not at cross purposes with scientific needs. The iterations shown here were accomplished in under a minute by our experienced artist. It is this flexibility and rapid iteration that enables artistic discovery. The key is to remove the barriers to achieving this creative state so that artists can use their artistic skills built up over a lifetime of experience to add their voices and visions to the multidisciplinary efforts to wrangle increasingly large and complex scientific data.
Science and art have a common thread in that many scientific breakthroughs, like artistic breakthroughs, happen when intense thought, contemplation, and experimentation meet the subconscious and the accidental, underscoring the need for artistic expression in our society and for science.





\subsubsection{Limitations}

We have described some feedback from artists on the occasion of their first
introduction to the interface, and it is important to note that this is not a
system to be mastered in a single session. The scope of the vocabulary feels
limitless, which is a good thing; artists are used to starting with limitless
possibilities and narrowing to the essential. However, the possibilities here
are even more expansive than sitting down in front of a canvas and paints; it is
more like walking into a new studio with paint, clay, printing press, video
software, and more. Artists tell us that they know they will need more time to
experiment before they can take full advantage of the tool. Similar to artists
learning a new medium, we fully expect that as artists spend more time with the
interface, the resulting variety in visualizations will expand as well.




We also wish to be clear that for all of the benefits for accessibility presented in this interface, it does not address the major challenge of data wrangling.  For the data pictured in this paper, we have relied upon ParaView to accomplish that task.  By structuring the design interface and rendering engine as a modular system that can connect to existing tools like ParaView via network sockets, we are able to read data from this tool that scientists already use for analysis and have a communication strategy that can be mimicked with other scientist-facing data analysis tools.  However, this does not address the problem that, for many actively studied datasets, the initial task bringing data into such tools can be a daunting challenge.

%% file: TEXT_Sections/7._Conclusion.tex
\section{Conclusions and Future Work}

Throughout the construction of this artist-focused interface for creating visualizations of 3D multivariate data, we have broken apart the conceptual components of existing visualization software and reassembled them into a new whole, leveraging artistic metaphor and language to present the data visualization design process in a new way.
Like prior work in building visualization interfaces that support artists in their creative methodologies, the interface emphasizes rapid exploration of design alternatives, and by building the interface on top of the Artifact-Based Rendering engine, the result also enables artists to bring color, texture, line, and form from their existing art practices to the world of data-intensive science.  
Our future work includes continuing to explore new visual encoding styles, which we plan to add to the interface as new plates.  We are also excited by the potential of tangible user interfaces that might create an even tighter connection and seamless workflow between the physical task of creating in the studio and incorporating those creations into data-driven 3D renderings.

%% file: MAIN.bbl
\begin{thebibliography}{10}

\bibitem{materialDesign}
Material design icons.
\newblock https://material.io/resources/icons/, Accessed 1st February, 2020.

\bibitem{greg-ibm-data-explorer}
G.~Abram and L.~Treinish.
\newblock An extended data-flow architecture for data analysis and
  visualization.
\newblock {\em ACM SIGGRAPH Computer Graphics}, 29(2):17--21, 1995.

\bibitem{ahrens2005paraview}
J.~Ahrens, B.~Geveci, and C.~Law.
\newblock Paraview: An end-user tool for large data visualization.
\newblock {\em The visualization handbook}, 717, 2005.

\bibitem{almeraj2009pencilLines}
Z.~AlMeraj, B.~Wyvill, T.~Isenberg, A.~A. Gooch, and R.~Guy.
\newblock Automatically mimicking unique hand-drawn pencil lines.
\newblock {\em Computers \& Graphics}, 33(4):496--508, 2009.

\bibitem{bach2017emerging}
B.~Bach, N.~H. Riche, S.~Carpendale, and H.~Pfister.
\newblock The emerging genre of data comics.
\newblock {\em IEEE computer graphics and applications}, 37(3):6--13, 2017.

\bibitem{bruckner2005volumeshop}
S.~Bruckner and M.~E. Groller.
\newblock {\em Volumeshop: An interactive system for direct volume
  illustration}.
\newblock IEEE, 2005.

\bibitem{buxton2010sketching}
B.~Buxton.
\newblock {\em Sketching user experiences: getting the design right and the
  right design}.
\newblock Morgan kaufmann, 2010.

\bibitem{FunctionalArt}
A.~Cairo.
\newblock {\em The Functional Art:An introduction to information graphics and
  visualization}.
\newblock New Riders, Berkeley, CA, 2013.

\bibitem{HPV:VisIt}
H.~Childs, E.~Brugger, B.~Whitlock, J.~Meredith, S.~Ahern, D.~Pugmire,
  K.~Biagas, M.~Miller, C.~Harrison, G.~H. Weber, H.~Krishnan, T.~Fogal,
  A.~Sanderson, C.~Garth, E.~W. Bethel, D.~Camp, O.~R\"{u}bel, M.~Durant, J.~M.
  Favre, and P.~Navr\'{a}til.
\newblock {VisIt: An End-User Tool For Visualizing and Analyzing Very Large
  Data}.
\newblock In {\em {High Performance Visualization--Enabling Extreme-Scale
  Scientific Insight}}, pp. 357--372. Oct 2012.

\bibitem{cox1988using}
D.~J. Cox.
\newblock Using the supercomputer to visualize higher dimensions: An artist's
  contribution to scientific visualization.
\newblock {\em Leonardo}, 21(3):233--242, 1988.

\bibitem{jim-gray-fourth-paradigm}
J.~Gray.
\newblock Jim {G}ray on e{S}cience: A transformed scientific method (2007).

\bibitem{guo2011wysiwyg}
H.~Guo, N.~Mao, and X.~Yuan.
\newblock Wysiwyg (what you see is what you get) volume visualization.
\newblock {\em IEEE Transactions on Visualization and Computer Graphics},
  17(12):2106--2114, 2011.

\bibitem{Houser}
H.~Houser.
\newblock {\em Infowhelm: Environmental Art and Literature in an Age of Data}.
\newblock Columbia University Press, New York, NY, 2020.

\bibitem{isenberg2008interactive}
T.~Isenberg, M.~H. Everts, J.~Grubert, and S.~Carpendale.
\newblock Interactive exploratory visualization of 2d vector fields.
\newblock In {\em Computer Graphics Forum}, vol.~27, pp. 983--990. Wiley Online
  Library, 2008.

\bibitem{johnson2019artifact}
S.~Johnson, F.~Samsel, G.~Abram, D.~Olson, A.~J. Solis, B.~Herman, P.~J.
  Wolfram, C.~Lenglet, and D.~F. Keefe.
\newblock {Artifact-{B}ased {R}endering: Harnessing Natural and Traditional
  Visual Media for More Expressive and Engaging {3D} Visualizations}.
\newblock {\em IEEE transactions on visualization and computer graphics},
  26(1):492--502, 2019.

\bibitem{keefe2008scientific}
D.~F. Keefe, D.~Acevedo, J.~Miles, F.~Drury, S.~M. Swartz, and D.~H. Laidlaw.
\newblock Scientific sketching for collaborative vr visualization design.
\newblock {\em IEEE Transactions on Visualization and Computer Graphics},
  14(4):835--847, 2008.

\bibitem{keefe2005artistic}
D.~F. Keefe, D.~B. Karelitz, E.~L. Vote, and D.~H. Laidlaw.
\newblock Artistic collaboration in designing vr visualizations.
\newblock {\em IEEE Computer Graphics and Applications}, 25(2):18--23, 2005.

\bibitem{lu2003stipple}
A.~Lu, C.~J. Morris, J.~Taylor, D.~S. Ebert, C.~Hansen, P.~Rheingans, and
  M.~Hartner.
\newblock Illustrative interactive stipple rendering.
\newblock {\em IEEE Transactions on Visualization and Computer Graphics},
  9(2):127--138, 2003.

\bibitem{luke2002semotus}
E.~J. Luke and C.~D. Hansen.
\newblock {\em Semotus visum: a flexible remote visualization framework}.
\newblock IEEE, 2002.

\bibitem{lum2002nprVolRend}
E.~B. Lum and K.-L. Ma.
\newblock Hardware-accelerated parallel non-photorealistic volume rendering.
\newblock In {\em Proceedings of the 2nd international symposium on
  Non-photorealistic animation and rendering}, pp. 67--ff, 2002.

\bibitem{Mielbach}
N.~Mielbach.
\newblock Straight talk with {N}athalie {M}iebach, 2013.

\bibitem{pasternak2017blockly}
E.~Pasternak, R.~Fenichel, and A.~N. Marshall.
\newblock Tips for creating a block language with {B}lockly.
\newblock In {\em 2017 IEEE Blocks and Beyond Workshop (B\&B)}, pp. 21--24.
  IEEE, 2017.

\bibitem{petersen2019antarctic}
M.~R. Petersen, X.~S. Asay-Davis, A.~S. Berres, Q.~Chen, N.~Feige, M.~J.
  Hoffman, D.~W. Jacobsen, P.~W. Jones, M.~E. Maltrud, S.~F. Price, et~al.
\newblock An evaluation of the ocean and sea ice climate of e3sm using mpas and
  interannual core-ii forcing.
\newblock {\em Journal of Advances in Modeling Earth Systems},
  11(5):1438--1458, 2019.

\bibitem{Petersen_ea15om}
M.~R. Petersen, D.~W. Jacobsen, T.~D. Ringler, M.~W. Hecht, and M.~E. Maltrud.
\newblock Evaluation of the arbitrary {Lagrangian-Eulerian} vertical coordinate
  method in the {MPAS-Ocean} model.
\newblock {\em Ocean Modelling}, 86(0):93 -- 113, 2015. doi: {{%
10\hspace{.1pt}\discretionary{.}{%
}{.}\hspace{.4pt}1016\discretionary{/}{%
}{/}j\hspace{.1pt}\discretionary{.}{%
}{.}\hspace{.4pt}ocemod\hspace{.1pt}\discretionary{.}{%
}{.}\hspace{.4pt}2014\hspace{.1pt}\discretionary{.}{%
}{.}\hspace{.4pt}12\hspace{.1pt}\discretionary{.}{%
}{.}\hspace{.4pt}004}}


\bibitem{polli2005atmospherics}
A.~Polli.
\newblock Atmospherics/weather works: A spatialized meteorological data
  sonification project.
\newblock {\em Leonardo}, 38(1):31--36, 2005.

\bibitem{Processing}
C.~Reas and B.~Fry.
\newblock {\em Processing: A Programming Handbook for Visual Designers and
  Artists}.
\newblock MIT Press, Cambridge, MA, 2014.

\bibitem{resnick2009scratch}
M.~Resnick, J.~Maloney, A.~Monroy-Hern{\'a}ndez, N.~Rusk, E.~Eastmond,
  K.~Brennan, A.~Millner, E.~Rosenbaum, J.~Silver, B.~Silverman, et~al.
\newblock Scratch: programming for all.
\newblock {\em Communications of the ACM}, 52(11):60--67, 2009.

\bibitem{Ringler_ea13om}
T.~Ringler, M.~Petersen, R.~Higdon, D.~Jacobsen, P.~Jones, and M.~Maltrud.
\newblock A multi-resolution approach to global ocean modeling.
\newblock {\em Ocean Modelling}, 69(0):211--232, 2013. doi: {{%
10\hspace{.1pt}\discretionary{.}{%
}{.}\hspace{.4pt}1016\discretionary{/}{%
}{/}j\hspace{.1pt}\discretionary{.}{%
}{.}\hspace{.4pt}ocemod\hspace{.1pt}\discretionary{.}{%
}{.}\hspace{.4pt}2013\hspace{.1pt}\discretionary{.}{%
}{.}\hspace{.4pt}04\hspace{.1pt}\discretionary{.}{%
}{.}\hspace{.4pt}010}}


\bibitem{Samsel2013ArtSciTech}
F.~Samsel.
\newblock {A}rt-{S}ci-{T}ech: {E}xamining the {S}pectrum.
\newblock {\em IEEE VIS, VISAP}, 2013.

\bibitem{SculptingVisLibrary}
F.~Samsel, G.~Abram, S.~Johnson, B.~Herman, and D.~F. Keefe.
\newblock The {S}culpting {V}is {L}ibrary.
\newblock http://sculpting-vis.tacc.utexas.edu/library.

\bibitem{samsel-visap-last-year}
F.~Samsel, A.~Bares, S.~Johnson, and D.~F. Keefe.
\newblock {Scientific Visualization: Enriching Vocabulary via the Human Hand}.
\newblock {IEEE} {VIS} Arts Program, 2019.

\bibitem{samsel2018colormoves}
F.~Samsel, S.~Klaassen, and D.~H. Rogers.
\newblock Colormoves: Real-time interactive colormap construction for
  scientific visualization.
\newblock {\em IEEE computer graphics and applications}, 38(1):20--29, 2018.

\bibitem{schroeder2010drawing}
D.~Schroeder, D.~M. Coffey, and D.~F. Keefe.
\newblock Drawing with the flow: a sketch-based interface for illustrative
  visualization of 2d vector fields.
\newblock In {\em SBM}, pp. 49--56. Citeseer, 2010.

\bibitem{schroeder2015visualization}
D.~Schroeder and D.~F. Keefe.
\newblock Visualization-by-sketching: An artist's interface for creating
  multivariate time-varying data visualizations.
\newblock {\em IEEE transactions on visualization and computer graphics},
  22(1):877--885, 2015.

\bibitem{sengal2015glacier}
A.~Segal.
\newblock Grewingk glacier.
\newblock https://www.adriensegal.com/grewingk-glacier, 2015.
\newblock Accessed March 2019.

\bibitem{shneiderman2000creating}
B.~Shneiderman.
\newblock Creating creativity: user interfaces for supporting innovation.
\newblock {\em ACM Transactions on Computer-Human Interaction (TOCHI)},
  7(1):114--138, 2000.

\bibitem{shneiderman2004designing}
B.~Shneiderman.
\newblock Designing for fun: how can we design user interfaces to be more fun?
\newblock {\em interactions}, 11(5):48--50, 2004.

\bibitem{singh2018orbacles}
G.~Singh.
\newblock Wearing multiple hats.
\newblock {\em IEEE Computer Graphics and Applications}, 38(4):6--8, 2018.

\bibitem{swackhamer2017weather}
M.~Swackhamer, A.~J. Johnson, D.~Keefe, S.~Johnson, R.~Altheimer, and
  A.~Wittkamper.
\newblock Weather report: Structuring data experience in the built environment.
\newblock {\em Proceedings of Architectural Research Centers Consortium}, pp.
  102--111, 2017.

\bibitem{terry2002sideViews}
M.~Terry and E.~D. Mynatt.
\newblock Side views: persistent, on-demand previews for open-ended tasks.
\newblock In {\em Proceedings of the 15th annual ACM symposium on User
  interface software and technology}, pp. 71--80, 2002.

\bibitem{Carpendale-2018-art-connection}
A.~Thudt.
\newblock Visualizations for personal reflection and expression.
\newblock 2018.

\bibitem{Viegas2007}
F.~Viegas and M.~Wattenberg.
\newblock {\em Artistic Data Visualization: Beyond Visual Analytics}, pp.
  182--191.
\newblock Springer Berlin Heidelberg, Berlin, Heidelberg, 2007. doi: {{%
10\hspace{.1pt}\discretionary{.}{%
}{.}\hspace{.4pt}1007\discretionary{/}{%
}{/}978\discretionary{%
}{-}{-}3\discretionary{%
}{-}{-}540\discretionary{%
}{-}{-}73257\discretionary{%
}{-}{-}0\_21}}


\bibitem{wang2019emotional}
Y.~Wang, A.~Segal, R.~Klatzky, D.~F. Keefe, P.~Isenberg, J.~Hurtienne,
  E.~Hornecker, T.~Dwyer, and S.~Barrass.
\newblock An emotional response to the value of visualization.
\newblock {\em IEEE computer graphics and applications}, 39(5):8--17, 2019.

\bibitem{Ware_2012}
C.~Ware.
\newblock {\em Information Visualization: Perception for Design}.
\newblock Morgan Kaufmann Publishers, San Francisco, CA, 2012.

\bibitem{west2015dataremix}
R.~West, R.~Malina, J.~Lewis, S.~Gresham-Lancaster, A.~Borsani, B.~Merlo, and
  L.~Wang.
\newblock Dataremix: Designing the datamade.
\newblock {\em Leonardo}, 2015.

\bibitem{wolfram2015diagnosing}
P.~J. Wolfram, T.~D. Ringler, M.~E. Maltrud, D.~W. Jacobsen, and M.~R.
  Petersen.
\newblock Diagnosing isopycnal diffusivity in an eddying, idealized midlatitude
  ocean basin via lagrangian, in situ, global, high-performance particle
  tracking (light).
\newblock {\em Journal of Physical Oceanography}, 45(8):2114--2133, 2015.

\bibitem{zhu2013d3}
N.~Q. Zhu.
\newblock {\em Data visualization with D3. js cookbook}.
\newblock Packt Publishing Ltd, 2013.

\end{thebibliography}
